\documentclass[twocolumn,showpacs,preprintnumbers,amsmath,amssymb]{revtex4-1}

\usepackage{graphics}
\usepackage{graphicx}
\usepackage{amsmath}
\usepackage{amsfonts}
\usepackage{amssymb}
\usepackage{pstricks}

\newcommand{\ex}{\hat{\vec{e}}_x}

\renewcommand{\vec}[1]{\mathbf{#1}}

\begin{document}

\title{The probability distribution of a trapped Brownian particle in plane shear flows}

\author{Jochen Bammert and Walter Zimmermann}

\affiliation{ Theoretische Physik I, Universit\"at Bayreuth, 95440 Bayreuth, Germany}

\date{Received: \today / Revised version: \today}

\begin{abstract}
We investigate the statistical properties of an over-damped Brownian particle
that is trapped by a harmonic potential and simultaneously exposed to
a linear shear flow or to a plane
Poiseuille flow. Its probability distribution is determined via the
corresponding Smoluchowski equation, which is solved analytically 
for a linear shear flow.
In the case of a plane Poiseuille flow,
analytical approximations for the distribution are obtained by a 
perturbation analysis and are substantiated by 
numerical results. There is a good agreement
between the two approaches
for a wide range of parameters.
\end{abstract} 


\maketitle

The dynamics of Brownian particles in fluids is of central importance 
in many areas of science \cite{Dhont:96,Berg:1993}.
There is a profound understanding of Brownian motion in quiescent
fluids, but the situation is different for
particles in flows. 
Several statistical properties 
of free Brownian particles in open flows
have been investigated,
for instance, in terms of the corresponding Langevin and
Fokker-Planck equations \cite{vdVen:1980.1}.
Recent works also consider inertia effects on diffusion in shear
flows \cite{Drossinos:05,Drossinos:2009.1} and the relation
between the Gaussian nature of noise and time-reversibility 
in driven systems \cite{Rubi:2007.1}.
It has also been
shown that shear flow causes, compared to a quiescent
fluid, additional correlations
of the particle's velocity and positional fluctuations
\cite{Bedeaux:1995.1,Brady:04,Drossinos:05}. 
However, the detection of  statistical 
properties of free particles in
a flow is an intricate issue.

To overcome part of these problems, 
three recent works focused 
on the dynamics of
Brownian particles trapped by harmonic potentials and
exposed to shear flows \cite{Ziehl:2009.1,Holzer:2009.1,Bammert:2010.1}.
It was shown that the shear-induced correlations between
the positional fluctuations of a captured particle along orthogonal
directions are essentially the same as in the free-particle case \cite{Holzer:2009.1}.
Furthermore, a  surprisingly good agreement was
found between the predictions and the measurements of these correlations
 \cite{Ziehl:2009.1}.

In Ref.~\cite{Holzer:2009.1}
the probability distribution of a trapped Brownian particle
in shear flows was obtained
by simulations of the Langevin equation.
In this brief report the related Smoluchowski equation 
for the probability distribution is presented and for
a linear shear flow an exact analytical solution is given.
For a plane Poiseuille flow we determine approximate analytical solutions,
which are in good agreement for a wide parameter range
with numerical solutions of the Smoluchowski equation and with
simulations of the Langevin equation as well.

We consider a Brownian particle trapped by a harmonic potential
with the spring constant $k$ at the origin
of the Cartesian coordinate frame $( {\bar x}, {\bar y}, {\bar z})$:
\begin{align}
\label{pot}
  V(\vec{\bar{r}})=\frac{k}{2} \vec{\bar{r}}^2\,.
\end{align}
The particle is exposed to a flow along the ${\bar x}$-direction
with a ${\bar y}$-dependent magnitude:
\begin{align} 
\label{flow}
\vec{u}(\vec{\bar{r}})=(a+b\bar{y}+c\bar{y}^2)\ex\,.
\end{align}
Since the Brownian motion perpendicular to the shear plane is
decoupled from the one in the shear plane, we consider 
the quasi two-dimensional problem with 
$\vec{\bar{r}}=(\bar{x},\bar{y})$ and
 $\bar{\nabla}=(\partial_{\bar{x}}, \partial_{\bar{y}})$.

For a plane Poiseuille flow the position of
the potential minimum can be different from the center of the flow.
In the shifted coordinate frame, 
$( {\tilde x}, {\tilde y})=  ( {\bar x}, {\bar y} + \tilde y_0)$, where
$\tilde{y}_0$ describes the $y$-coordinate of the potential minimum, the flow profile is given by 
$\vec{u}({\tilde{y}})=u_p(1-\tilde{y}^2/l^2)\ex$
with  the flow velocity $u_p$ at its center and
 the confining plane boundaries at $\tilde{y}=\pm l$.
If  a particle is trapped at $\tilde{y}_0 \not =0$,
one obtains with $\tilde{y}=\tilde{y}_0+\bar{y}$ the coefficients
$a= u_p (1-\tilde{y}_0^2/l^2)$, $b=-2u_p \tilde{y}_0/l^2$ and $c = -u_p/ l^2$ in Eq.~(\ref{flow}).
This work focuses on situations where the particle positions are 
sufficiently far away from the boundaries, so that hydrodynamic
interactions with the wall, as discussed in \cite{vdVen:1980.1} for instance,
can be neglected.

The particle dynamics are determined by thermal motion, 
potential forces and drag forces caused by the flow.
 The thermal motion is
characterized by the diffusion constant $D=k_BT/\zeta$, which is given
by the Einstein relation in terms of the temperature $T$, the Boltzmann constant $k_B$ and
the Stokes friction coefficient $\zeta=6 \pi \eta R$, where $\eta$ is the 
viscosity of the fluid and $R$ is the effective radius of the particle.
The external flow $\vec{u}$ is the origin of the Stokes drag force
$\zeta \vec{u}$ on the point-like particle, which is balanced by
the restoring force $-\bar{\nabla} V= -k \bar{\vec{r}}$.

The diffusion and the two deterministic forces drive
the probability current $\vec{j}(x,y,t)$
of the Smoluchowski equation (SE) for the particle's positional distribution function 
$\mathcal{P}(\bar{x},\bar{y},\bar{t})$ \cite{Risken:89}:
\begin{align}
 \label{conti}
\partial_{\bar t} \mathcal{P}  &= - \bar \nabla \vec{j}\,, \\
 \label{fick}
\vec{j}&=-D \bar{\nabla} \mathcal{P} + (\vec{u} - \frac{1}{\zeta}\nabla V) \mathcal{P}\,.
\end{align}

With the expressions in Eqs.~(\ref{pot}) - (\ref{fick}) 
the SE of a Brownian particle trapped in a harmonic potential and
exposed to a shear flow reads:
\begin{align}
\label{Smoluchowski_old}
\partial_{\bar t} \mathcal{P} =& D \bar{\nabla}^2 \mathcal{P} + \frac{D}{\delta^2} \left(2 +  \bar{x}\partial_{\bar{x}} 
+  \bar{y}\partial_{\bar{y}}\right) \mathcal{P}   \nonumber\\
&-(a+b \bar{y}+ c \bar{y}^2)  \partial_{\bar{x}} \mathcal{P}   \,.
\end{align}
The two spatial coordinates 
may be rescaled by the length $\delta=\sqrt{k_BT/k}$, $\bar{x} = \delta x$ and $\bar{y}=\delta y$, 
alike the time, ${\bar t} =\zeta/k\, t$. 
This results in the dimensionless SE
\begin{align}
\label{Smoluchowski}
\partial_t \mathcal{P} = \left[\nabla^2  + 2 + y\partial_y  
+(x-\alpha-\beta y-\gamma y^2)  \partial_x \right]\mathcal{P} \,,
\end{align}
with the parameters $\alpha=a \delta/D$, 
$\beta=b \delta^2/D$ and $\gamma= c \delta^3/D$ 
describing the flow profile and $y \in [(-l-\tilde{y}_0)/\delta,\, (l-\tilde{y}_0)/\delta]$. 
$\beta$ is the so-called Weissenberg number.
We note here, that a modified Smoluchowski equation including
inertia in shear flows is presented in \cite{Drossinos:05,Drossinos:2009.1}.
In comparison to these works, the presence of the potential (\ref{pot})
ensures a stationary solution $\mathcal{P}(x,y)$.

\begin{figure}
 \centering
\includegraphics[width=6cm]{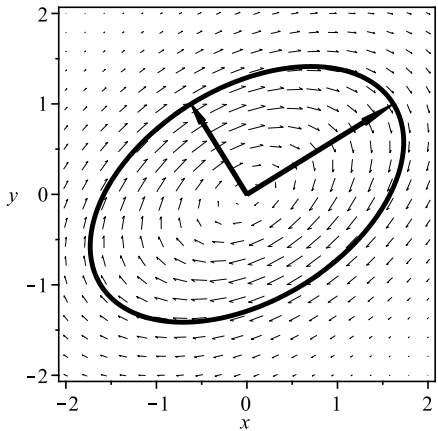}
 \vspace*{-4mm}
 \caption{An elliptical contour line of the distribution of a trapped particle 
in a linear shear flow is shown, 
cf. Eq.~(\ref{ansatzps})  with $\alpha=0, \beta=1$, as well as the vector field $\vec{j}(x,y)$.
The broad arrows indicate the two principal axes of the ellipse.
 \label{PDF_shear}}
\end{figure}

For a uniform flow, i.e.,  $\alpha \not =0$ and
$\beta=\gamma=0$, the static solution
of Eq.~(\ref{Smoluchowski}) is the shifted Boltzmann distribution
\begin{align}
 \label{pa0b0c0}
\mathcal{P}_n(x,y) = P_0 e^{-\frac{1}{2}x^2-\frac{1}{2}y^2+\alpha x}\,,
\end{align}
where $P_0=(2\pi e^{\alpha^2/2})^{-1}$ is determined
by  $ \int \int dx dy P_n(x,y)=1$.
A superposition of a uniform flow and
a linear shear flow, i.e., $\alpha \neq 0,\beta \neq 0$ and $\gamma=0$,
leads to the expected Gaussian distribution 
\begin{align}
\label{ansatzps}
 \mathcal{P}_{ab}(x,y)=P_0 e^{-(a_1x^2+ a_2y^2  + a_3xy + a_4x + a_5y)}\,,
\end{align}
with  the coefficients
\begin{subequations} \label{coeffps}
\begin{align}
 &a_1= \frac{2}{\beta ^2+4}\,,~~a_2= \frac{\beta ^2+2}{\beta ^2+4}\,,~~ a_3= \frac{-2 \beta }{\beta ^2+4}\,, \\
 &a_4= \frac{-4\alpha}{\beta ^2+4}\,,~~a_5= \frac{2\alpha \beta }{\beta ^2+4}\,,
\end{align}
\end{subequations}
and the norm $P_0~=~\left(\pi\,  \sqrt{\beta^2+4}\,
e^{2\alpha^2/(\beta^2+4)}   \right)^{-1}$.
In the case of a finite Weissenberg number $\beta$ the 
contour lines of the
probability distribution $\mathcal{P}_{ab}(x,y)$ 
are elliptical, as shown in Fig.~\ref{PDF_shear}.
The probability current $\vec{j}(x,y)$, indicated by the vector field
in the same figure, characterizes the mean particle motion.

The parameter $\alpha$, 
describing the contribution of the uniform flow,
causes essentially a shift of the distribution $\mathcal{P}_{ab}(x,y)$ in the $x$-direction.
In the following we consider the case $\alpha=0$, where $a_4=a_5=0$ and
the resulting distribution function is denoted by $\mathcal{P}_a(x,y)$.
Its elliptical contour lines can be characterized by two principal axes.
The ratio between their lengths as well as the angle between the major principal
axis and the flow direction are a function 
of the Weissenberg number $\beta$.
This was already discussed in Ref.~\cite{Holzer:2009.1},
where shear-induced corrections to the autocorrelations
as well as a cross-correlation between orthogonal particle fluctuations
in the shear plane were found.
These correlations can also be calculated in terms
of the probability distribution $\mathcal{P}_a(x,y)$ via 
the expression $\langle r_i r_j \rangle = \int \int dx dy \mathcal{P}_a(x,y)r_i r_j$:
\begin{align}
 \langle xx\rangle=  1+ \frac{\beta ^2}{2},\quad \langle xy\rangle=\langle yx\rangle= \frac{\beta }{2}, \quad \langle yy\rangle=1\,.
\end{align}
After rescaling to dimensional units, $x \to \bar x $, the
results given in Ref.~\cite{Holzer:2009.1} are recovered.

In a plane Poiseuille flow all three parameters 
$\alpha$, $\beta$ and $\gamma$ in the SE~(\ref{Smoluchowski})
may be non-zero and no exact analytical solution was found in this case.
Similar to \cite{vdVen:1980.1}, the
probability distribution is calculated perturbatively and compared with
numerical solutions of Eq.~(\ref{Smoluchowski}).

First we consider a Brownian particle which is trapped at the center
of a plane Poiseuille flow, with $\tilde{y}_0=0$. 
In this case of vanishing $\beta$, the parameter
$\alpha=u_p\delta\zeta/(k_BT)$ describes the ratio between the
drag force imposed by the flow 
and the potential force on the particle.
If $\alpha$ is fixed, the parameter
$\gamma=-\alpha(\delta/l)^2$ depends on the ratio between the
two characteristic lengths $l$ and $\delta$. Note, the
hydrodynamic interactions between the particle and the
walls are small in the range $\delta/l < 1/2$.

Our ansatz for the
 perturbation expansion of the solution
of Eq.~(\ref{Smoluchowski}) up to the second order in
 $\gamma$ reads
\begin{align}
\label{ansatzpert1}
 \mathcal{P}(x,y)=\mathcal{P}_{n}(x,y)~e^{\gamma f_1(x,y)+\gamma^2 g_1(x,y)}\,,
\end{align}
with the two polynomials
\begin{subequations}\label{p1exppoly}
\begin{align}
\label{p1exppoly1}
 f_1(x,y)&= b_1x+ b_2 y^2  + b_3xy^2\,, \\
 g_1(x,y)&=(c_1x +c_2x^2)(1+y^2) +c_3y^2 +c_4y^4\,.
\end{align}
\end{subequations}
The SE~(\ref{Smoluchowski}) may then
be rewritten: 
\begin{align}
\label{bestgl1}
\left(p_0(x,y)+\gamma p_1(x,y) + \gamma^2 p_2(x,y) \right) \mathcal{P}(x,y)=0\,.
\end{align}
Since $\gamma$ is an arbitrary, but small number, the polynomials
$p_{0,1,2}(x,y)$ in Eq. (\ref{bestgl1}) have to vanish separately.
According to  Eq. (\ref{ansatzps}) the condition $p_0(x,y)=0$ is 
automatically fulfilled. The second condition, 
$p_1(x,y)=0$, provides the first-order coefficients
\begin{align}
 \label{bk}
  &b_1=\frac{2}{3},\quad  b_2=\frac{-\alpha}{3},\quad b_3=\frac{1}{3}\,,
\end{align}
whereas the third condition, $p_2(x,y)=0$,
 determines the coefficients at $\mathcal{O}(\gamma^2)$:
\begin{align}
 \label{cl}
 &c_1=\frac{-2\alpha}{9},\quad c_2=\frac{1}{9},\quad  c_3=\frac{\alpha^2}{9}-\frac{1}{3},\quad c_4=\frac{-1}{18}\,.
\end{align}

With the potential minimum off the center of the Poiseuille flow, 
one has $\tilde{y}_0 \neq 0$, a finite
Weissenberg number  $\beta \neq 0$ and no longer $\pm y$ symmetry. 
Again we use an ansatz of the form:
\begin{align}
\label{ansatzpert2}
 \mathcal{P}(x,y)&=\mathcal{P}_{ab}(x,y)\, e^{\gamma f_2(x,y) + \gamma^2 g_2(x,y)}\,.
\end{align}
Due to the loss of the $\pm y$ symmetry,
the polynomial $f_2(x,y)$ has nine different contributions:
\begin{align}
\label{f2}
f_2(x,y)=&d_1x+d_2y+d_3(x-3\alpha)x^2+d_4(x-\alpha)y^2 \nonumber\\
 &+d_5(x-2\alpha)xy+d_6y^3\,.
\end{align}
The condition that the $\mathcal{O}(\gamma)$-terms in Eq.~(\ref{Smoluchowski})
have to vanish determines the coefficients in the ansatz (\ref{f2}).
With  $B=\beta^2+4$ and $E=2/(9B^3)$ they are given by
\begin{align}
 \label{ci}
 d_1&=4E\left( (12-\beta^2)B + 16\alpha^2 \beta^2\right),\nonumber\\
 d_2&=-2E\beta\left( (20+\beta^2)B + 8\alpha^2(\beta^2-4)\right),\nonumber\\
 d_3&=\frac{64}{3}E\beta^2\,,~~~
 d_4=2 E(3\beta^4-8\beta^2+48), \nonumber\\
 d_5&=16 E\beta (4-\beta^2),  \nonumber\\
 d_6&=\frac{-E\beta}{3}(5\beta^4+24\beta^2+144)\, . 
\end{align}
In the limit $\beta \to 0$ the coefficients (\ref{bk}) are recovered.
The polynomial $g_2(x,y)$ for the order $\gamma^2$
includes $14$ lengthy contributions, 
 which we do not list here.

In order to estimate the validity range of the 
approximations presented above,
we determine stationary solutions of the SE.~(\ref{Smoluchowski}) 
numerically. To this end a simple Jacobi-relaxation method, cf. Ref.~\cite{NumResF}, 
or a direct integration of the rescaled Eq.~(\ref{conti}) is sufficient.
The probability current is determined via:
\begin{align}
 \label{numerics2}
 \vec{j}&= -\nabla \mathcal{P} + \left[ \begin{pmatrix} \alpha+\beta y + \gamma y^2\\ 0  \end{pmatrix}
	-\begin{pmatrix} x\\ y  \end{pmatrix} \right]\mathcal{P}\,.
\end{align}
\begin{figure}[t]
\vspace{-2mm}
 \centering
  \includegraphics[width=7cm]{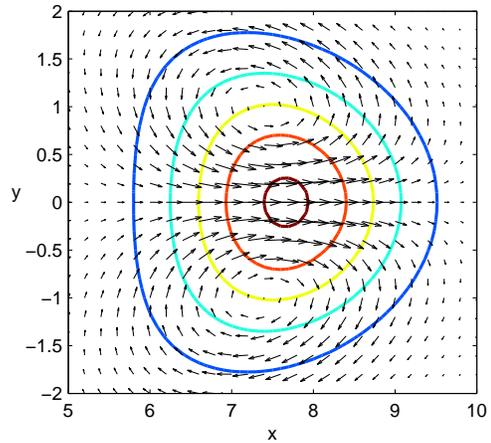}
\vspace{-2mm}
  \caption{Contour lines of the numerical solution of Eq.~(\ref{Smoluchowski}) as well as
 the vector field $\vec{j}(x,y)$ are shown for the case where 
 the potential minimum is at the center of
 a plane Poiseuille flow.
Parameters: $\alpha=8, ~\gamma=-\alpha(\delta/l)^2=-1/2$. 
  \label{PDF_poiscenter}}
\end{figure}

Fig.~\ref{PDF_poiscenter} shows the contour lines of the numerically obtained 
probability distribution $\mathcal{P}(x,y)$ in the case 
with the minimum of the trapping potential at the center of a plane Poiseuille flow.
This distribution has similarities with the parachute or bullet-like shape
of vesicles in a Poiseuille flow \cite{Skalak:1969.1,Gompper:2005.1,Misbah:2009.1}.
In comparison to similar numerical results, obtained via the related Langevin equation and 
presented in Ref.~\cite{Holzer:2009.1}, we additionally show
the probability current $\vec{j}(x,y)$.
As indicated in Fig.~\ref{PDF_poiscenter}, the vector field $\vec{j}({x,y})$
includes two counter-rotating vortices that are symmetric with respect to the $x$-axis.

\begin{figure}[t]
\vspace{-2mm}
  \centering
  \includegraphics[width=7cm]{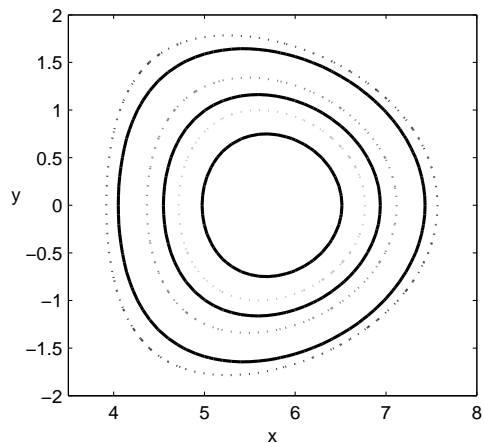}
\vspace{-2mm}
  \caption{Comparison of the numerically determined distribution
(solid lines) with the analytical approximations, given by Eq.~(\ref{ansatzpert1}) (dashed lines).
 The contour lines of both types
are plotted at different heights in order to distinguish them to compare
their shapes. Parameters: $\alpha=6$, $\gamma=-3/8$.
  \label{PDF_vgl1}}
\end{figure}

In Fig.~\ref{PDF_vgl1} contour lines 
of the numerically obtained probability 
distribution are compared  with those obtained
from the perturbative solution (\ref{ansatzpert1})
 for the parameters $\alpha=6$ and
$\gamma=-\alpha(\delta/l)^2=-3/8$. In spite of this
rather large
$|\gamma|$-value, beyond the expected validity range
of the perturbation expansion, the
differences between both solutions in Fig.~\ref{PDF_vgl1}
are surprisingly small.
As expected, for decreasing values of $|\gamma|$ these
differences become
even smaller, but the analytical formula (\ref{ansatzpert1}) 
may be useful for fitting
experimental data up to $|\gamma|\simeq 3/8$.

\begin{figure}[t]
  \centering
  \includegraphics[width=7cm]{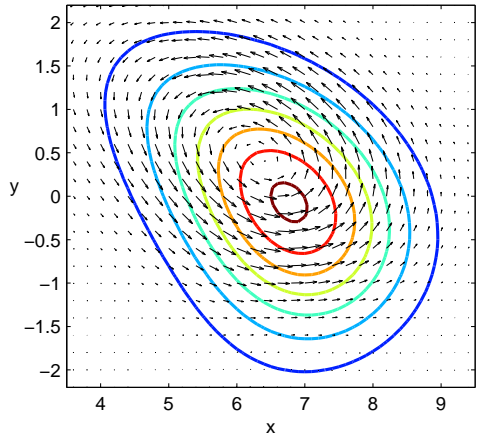}
\vspace{-2mm}
  \caption{Contour lines of $\mathcal{P}(x,y)$ and vector field 
$\vec{j}(x,y)$ for the case where the potential minimum is off the center of
 the Poiseuille flow. Parameters: $\alpha=7, \beta=-14/15, \gamma=-7/15$.
  \label{PDF_poisoff}}
\end{figure}

The symmetrical shape of the 
distribution is lost if the particle is trapped away from the center of the
plane Poiseuille flow.
With increasing values of $\tilde y_0$ and 
 $\beta$, the shape of the probability distribution deforms from a parachute or bullet 
towards an ellipse. Simultaneously one vortex
of the probability current is enhanced while the other one is weakened.
For the values $\tilde{y}_0=l/4$, $\delta/l=1/4$ and $\alpha=7$, with
 $\beta=-2\alpha/15$ and
$\gamma=-\alpha/15$, the contour lines of $\mathcal{P}(x,y)$ as well as the current $\vec{j}(x,y)$ 
are displayed in Fig.~\ref{PDF_poisoff}. In this case the lower vortex in $\vec{j}(x,y)$ has already vanished
and the distribution $\mathcal{P}(x,y)$ shares similarities with the 
slipper shape of vesicles in capillary flows \cite{Misbah:2009.2}. Again the results are in good agreement with the simulations in Ref.~\cite{Holzer:2009.1}.

In conclusion, we investigated the positional distribution  of 
a Brownian particle, which is trapped by a harmonic potential and simultaneously
exposed to different shear flows. A complete analytical solution
of the corresponding Smoluchowski equation is given 
in the case of a linear shear flow.  
For a plane Poiseuille flow, we presented
approximate analytical formulas,
which are in good agreement with numerical solutions for a wide range of parameters.
Some of our results confirm earlier ones
obtained in Ref.~\cite{Holzer:2009.1}
on the basis of simulations of the related Langevin equation.

Our predictions of the particle's probability distribution 
in a Poiseuille flow may be measured in an experiment
similar to that in Ref.~\cite{Ziehl:2009.1}. In this  work,
a micron-sized polystyrene bead 
was trapped by an optical tweezer and the 
fluctuating particle positions were recorded
by a high speed camera in a stroboscopic manner. For the positions of a 
trapped particle 
in Poiseuille flow one expects probability distributions
as predicted in this work. The presented
 analytical approximations for the particle distribution may be useful 
for fitting the experimental data.

Experiments with particles placed near the center of a Poiseuille flow,
where the flow velocity takes its maximum value, may require large laser intensities
in order to keep the particles trapped by the potential. 
This constraint reduces the flexibility for variations of
the typical length scale $\delta$
 of the particle's positional fluctuations. Since the
flow profile determines the shape of the particle distribution
and not the flow velocity near the 
potential minimum, one may reduce the drag force by moving the trap along with the flow.

The statistical properties of trapped Brownian particles, 
as investigated in \cite{Quake:2000.1,Ziehl:2009.1,Holzer:2009.1},
share similarities with those of tethered polymers
exposed to uniform flows \cite{Chu:1995.1,Brochard:1993.1,Rzehak:99.2,Rzehak:2000.1} 
or to shear flows \cite{Doyle:2000.1}. The related theoretical explorations 
are mainly based on Brownian dynamics simulations. 
An interesting question is,
how far can the analysis described here
be applied to tethered bead-spring models in shear flows?
Do such investigations exhibit temporal oscillations 
as found for instance for deterministic models \cite{Holzer:2006.1}? 

We thank M. Burgis for inspiring discussions.
This work was supported by DFG via the priority program on micro- and nanofluidics
SPP 1164, and by the Bayerisch-Franz\"osisches Hochschulzentrum.

\end{document}